\documentclass[conference]{IEEEtran}
\IEEEoverridecommandlockouts
\usepackage{cite}
\usepackage{amsmath,amssymb,amsfonts}
\usepackage{algorithmic}
\usepackage{graphicx}
\usepackage{textcomp}
\usepackage{xcolor}
\usepackage{siunitx}
\usepackage{subcaption}
\usepackage{hyperref}
\usepackage{dsfont}
\usepackage{bm}
\usepackage[numbers]{natbib}
\def\BibTeX{{\rm B\kern-.05em{\sc i\kern-.025em b}\kern-.08em
    T\kern-.1667em\lower.7ex\hbox{E}\kern-.125emX}}
\begin{document}

\newcommand{\minisection}[1]{\vspace{0.025in} \noindent {\bf #1}\ }

\title{Technical Challenges of Deploying Reinforcement Learning Agents for Game Testing in AAA Games\\
}

\author{

Jonas Gillberg$^{1}$,
Joakim Bergdahl$^2$,
Alessandro Sestini$^{2}$,
Andrew Eakins$^{1}$,
Linus Gisslén$^2$
\\
\textit{$^1$Electronic Arts (EA)}, 
\textit{$^2$SEED - Electronic Arts (EA)},  \\
jonas.gillberg@dice.se, 
aeakins@europe.ea.com, 
\{jbergdahl, asestini, lgisslen\}@ea.com \\
}


\IEEEoverridecommandlockouts

\IEEEpubid{\makebox[\columnwidth]{979-8-3503-2277-4/23/\$31.00~\copyright2023 IEEE \hfill} 
\hspace{\columnsep}\makebox[\columnwidth]{ }}

\maketitle

\IEEEpubidadjcol

\begin{abstract}
Going from research to production, especially for large and complex software systems, is fundamentally a hard problem. In large-scale game production, one of the main reasons is that the development environment can be very different from the final product. 
In this technical paper we describe an effort to add an experimental reinforcement learning system to an existing automated game testing solution based on scripted bots in order to increase its capacity. We report on how this reinforcement learning system was integrated with the aim to increase test coverage similar to \cite{bergdahl2020augmenting} in a set of AAA games including \textit{Battlefield 2042} and \textit{Dead Space (2023)}. The aim of this technical paper is to show a use-case of leveraging reinforcement learning in game production and cover some of the largest time sinks anyone who wants to make the same journey for their game may encounter. 
Furthermore, to help the game industry to adopt this technology faster, we propose a few research directions that we believe will be valuable and necessary for making machine learning, and especially reinforcement learning, an effective tool in game production.
\end{abstract}

\begin{IEEEkeywords}
game testing, reinforcement learning, scripted bots, AAA games, game production
\end{IEEEkeywords}

\section{Introduction}
\label{sec:introduction}

Automated testing is already a requirement in most modern, large-scale games. A case study is \textit{Battlefield V}, which requires testing of 601 different features amounting to around 0.5M hours of testing if done manually. This corresponds to~$\approx300$ work-years and with every game in the franchise, this number is growing. 
 Automated testing can cover many test cases, but the capacity of the scripted solution is already reaching its limit and an increasing number of cases require more sophisticated techniques for testing the game in depth. Common test solutions relying on script-based bots have several drawbacks: the bots do not learn from interactions, domain specific scripting expertise is required, and some use-cases are hard or even impossible to script for, to name a few.
 Recently, a promising research direction in game testing has evolved: it has been shown that agents trained via Reinforcement Learning (RL) can bring value to automated game testing~\cite{bergdahl2020augmenting, gordillo2021improving, sestini2022ccpt}. RL has the capacity to leverage exploration to increase test coverage, find exploits in gameplay, and detect imbalances (ranging from level design, and character design, to weapon design). Compared to classical solutions, RL allows for flexibility when introducing new game features as agents can easily be retrained instead of manually re-scripted. Another benefit of RL is that instead of requiring hand-authoring behaviors, they can be generated in a more high-level abstraction through reward functions which often are more intuitive: instead of scripting an agent to solve a task, the user can simply give it rewards relevant to the task and the agent may figure out how to solve it in an emergent fashion. Compared to many classical solutions, RL does not require a navigation mesh for pathfinding as the agent is controlled through the same controller inputs a player experiences. By not being bound to the navigation mesh, the agent enjoys a higher degree of freedom leading to behaviors not attainable using a scripted solution, which is useful for exploit discovery, for instance.
 
 As straightforward as this approach might seem, there are many hurdles to overcome before RL can be applied to games, especially when combined with existing automated systems that depend on scripting. Current RL research is commonly pursued in simple environments that do not cover the complexity and scope found in AAA game production. Fundamental research is naturally needed to push the envelope, and toy problems and simplified environments certainly play an important role in this. However, it also means that bringing current research into AAA games in a shippable form will cause unforeseen challenges to overcome. The use of RL agents alone should not be seen as means of replacing scripted bots altogether, but rather complement scripted bots in a symbiotic fashion where they fall short.

In this paper, we describe an effort to augment scripted-only automated game testing using RL, and all the technical and practical challenges that emerge from working with games in production. We first present current industry-standard test methods based on bot scripting, then we highlight a few real-world AAA games that are relevant to this study where classical test solutions are impractical, followed by a description of how RL can be used to mitigate this. We present lessons learned from applying RL to games in production and list challenges that need to be overcome when doing so. Finally, we outline related work on this topic and then conclude with a discussion and future work directions.

\begin{figure*}        
    \centering				
    \includegraphics[width=0.72\textwidth]{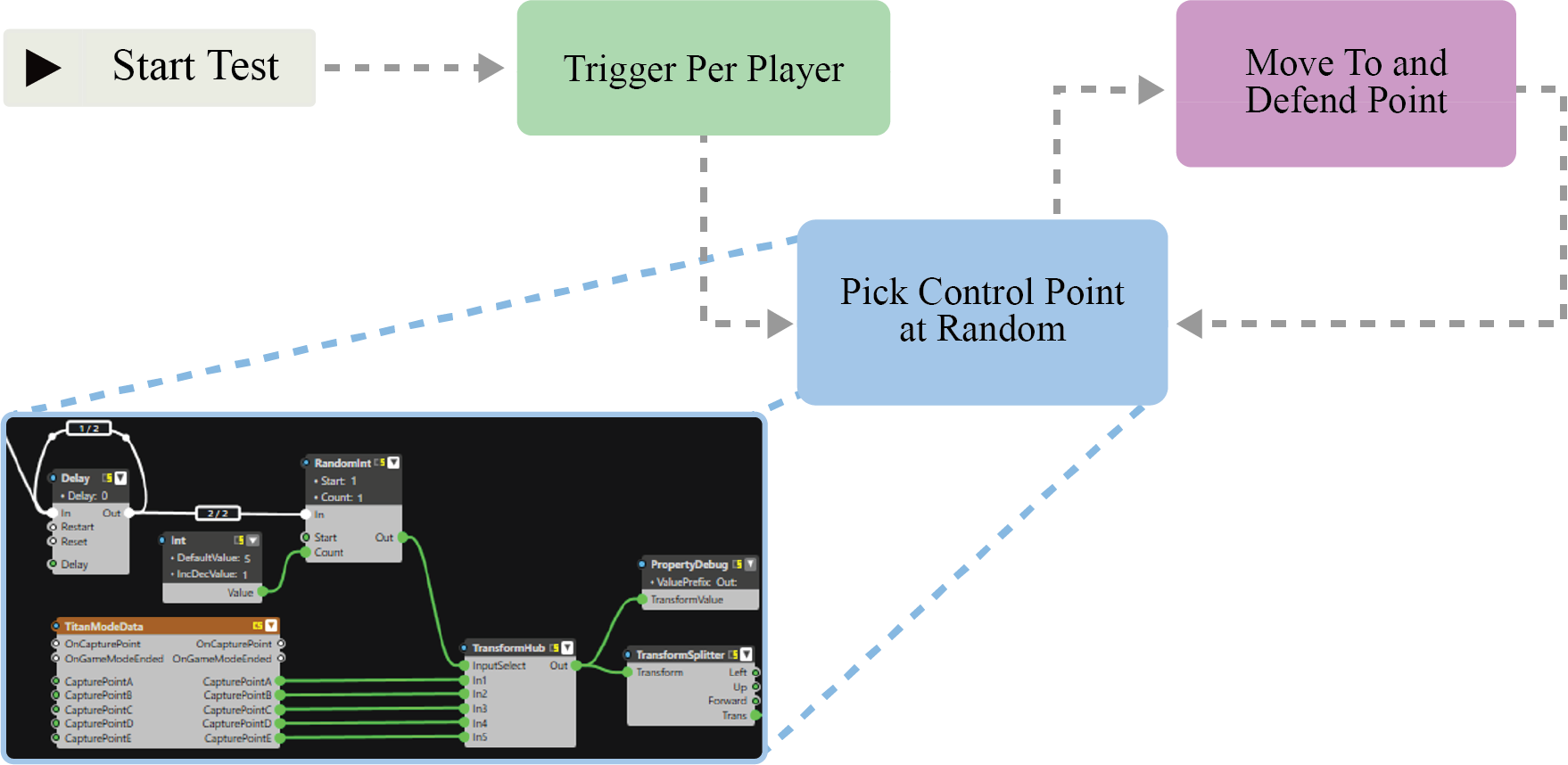}
    \caption{Visual scripting system for creating different behaviours. In this example, bots are tasked to defend capture points randomly selected from a predefined set. At a random time interval, a new point is sampled and selected.}
    \label{fig:schematics}	
\end{figure*}

\section{Scripted Bots for Game Testing}
\begin{figure}[h!]           
     \centering				
     \includegraphics[width=0.8\columnwidth]{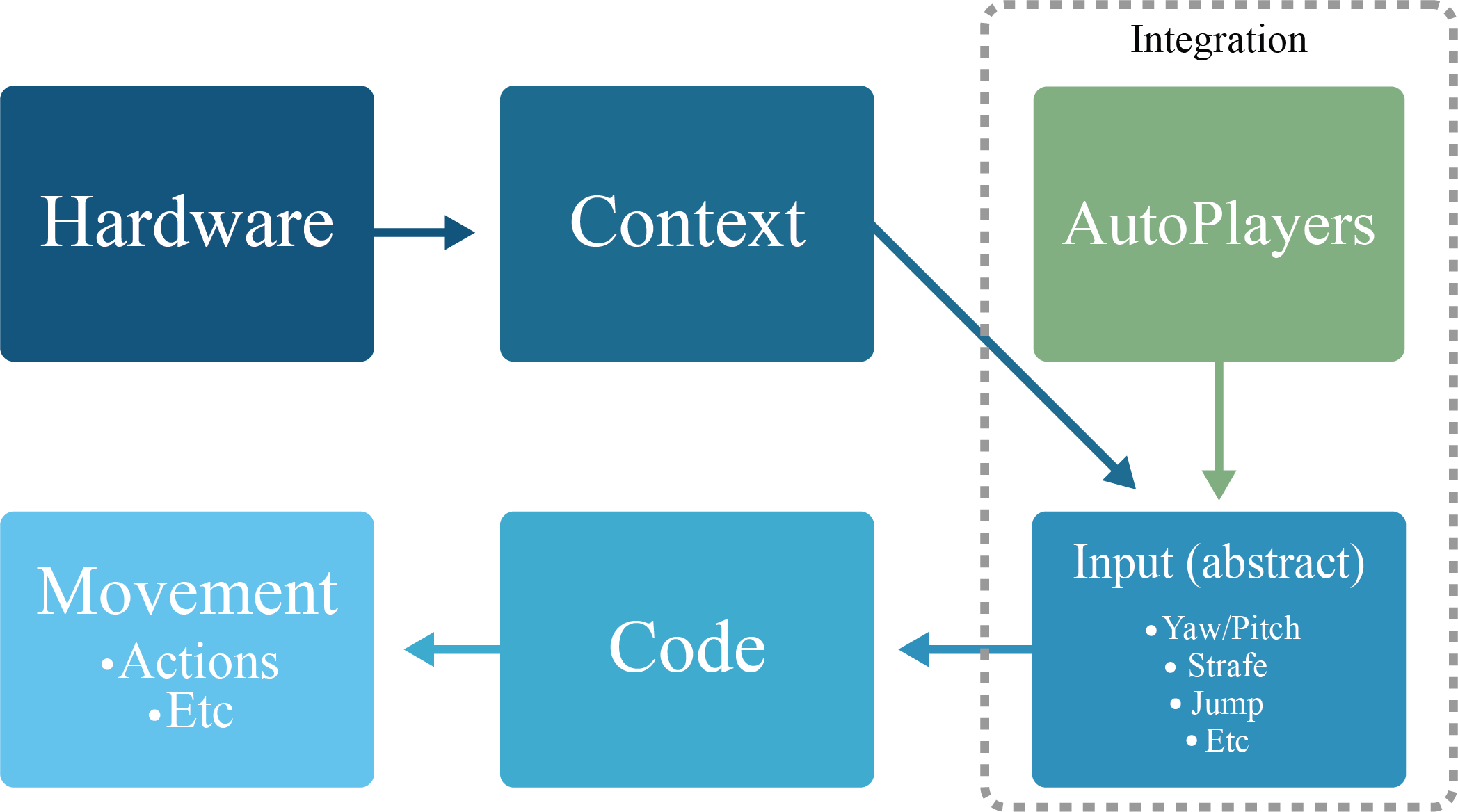}
     \caption{Overview of the integration of \textit{AutoPlayers} into the input handling system pre-existing in the game.}
     \label{fig:autoplayers_overview}	
 \end{figure}

\textit{AutoPlayers} is an in-house developed scripted bot system that is used for several internal games at \textit{Anonymized}, created to address the growing scale of modern games and their corresponding testing needs. In \textit{AutoPlayers}, bots control game characters in a way very similar to how a player would in an effort to generate test data that is as player representative as possible. The AI itself outputs action and movement intentions which are then translated into game specific inputs. In many cases, the most useful translation is to game or engine specific action representations (see Figure \ref{fig:autoplayers_overview}), rather than hardware inputs, meaning context specific translations are automatically handled by the game. An alternative approach that is sometimes applied is to perform contextual lookups of which corresponding hardware key currently would result in the desired action and feed that key into the system.

Scripted bots are especially useful in the many scenarios where a high degree of control is desired. The \textit{AutoPlayers} bots are controlled through the scripting of high level intentions called \textit{Objectives}. Objectives can either be generated by code or provided from scripts as can be seen in the visual scripting example in Figure~\ref{fig:schematics}.
Objectives control where the \textit{AutoPlayers} go, but objective parameters also control how they get there. If they should engage their enemies, remain passive, be invincible etc. Illustrating the typical granularity of these objectives, recurring type examples include: \textit{Move}, \textit{Defend}, \textit{Interact}, \textit{Action}, \textit{Attack}, \textit{Follow} and more high level ones like \textit{Seek And Destroy}, among others.


When an \textit{AutoPlayers} bot engages in combat with an enemy it will move and attack by relying on a predefined set of rules in code, based on metrics like enemy distance, effective weapon range among others. This bundled autonomous behavior concept is one of the key reasons setting up tests with \textit{AutoPlayers} is straight forward as only high level goals need to be specified and the details of how to accomplish them is left to lower level systems.

Of these systems, \textit{locomotion} is one of the most important. Locomotion is the system that calculates the appropriate movement controls to move the bot along a path of waypoints, typically produced as the result of a query to the navigation system. In the \textit{AutoPlayers} case, different versions of the locomotion system are used depending on whether the bot is, for example, moving on foot, piloting a helicopter or driving a tank.

One of the main challenges when developing test bots is that they need to always function throughout the games development cycle, long before the game is stable and the feature set has stabilized. As such, there is a constant trade-off between \textit{properly} testing features while at the same time not wasting development time on building bespoke test implementations for a feature that might go through major changes or be cut entirely.

\section{Use-cases and Environments}
\label{sec:environments}
In this section, we describe the games considered for this study, their properties, and how test solutions based on scripted bots may be impractical. Further, we define a practical recipe for relaxing the constraints seen in the full production games to make the process of training RL agents in them easier and faster by the means of \textit{test range} and \textit{prototype} environments.

\subsection{Production environments}
\label{sec:production_environment}
The game engine used for both of the production environments is an in-house game engine called \textit{Frostbite}. Both of these games contained preexisting integrations of \textit{AutoPlayers}. Here we describe the different environment variants that were used to evaluate our solution, combining \textit{AutoPlayers} and RL.

\minisection{Battlefield 2042.}
\label{sec:helicopters_kingston}
\textit{Battlefield 2042} is a team based, large-scale, multiplayer oriented first-person shooter game. This title has a multi-modal gameplay structure where the player tries to defeat enemies either as on-the-ground soldiers, in ground vehicles, or in aircrafts. The first use-case explored is helicopter navigation, shown in Figure~\ref{fig:kingston_helicopter}. In order to validate and test that helicopters can be successfully piloted to reach all intended locations in the game world and that they do not get stuck, we seek to find a way of controlling them through controller inputs. In pre-production, helicopters were very difficult to steer even for expert humans, leading to challenges when scripting control for them through the use of \textit{AutoPlayers}. This is further exacerbated by the continuously changing parameters governing flight dynamics during development and the lack of time to propery rewrite the control solution to compensate for this. To manage this ever-changing control problem, one solution is to train RL agents that automatically learn to control and navigate the vehicle through interactions with the game environment.


\minisection{Dead Space (2023).}
\textit{Dead Space} is a slow-paced, singleplayer, third-person shooter game with a sci-fi horror setting. An interesting characteristic of this title is that in some gameplay sections the player is tasked to navigate in \textit{zero gravity}, illustrated in Figure \ref{fig:dead_space_production}. In these sections, the player experiences $6$ degrees of freedom of movement, compared to the $3$ degrees in other areas where the player is bound to the floor. This combined with the physics properties of the game like inertia makes scripted control difficult and compensating for these may be non-trivial.

\subsection{Test range environments}
\begin{figure*}[t]
    \centering
    \begin{subfigure}[b]{0.32\textwidth}
        \centering
        \includegraphics[width=\textwidth]{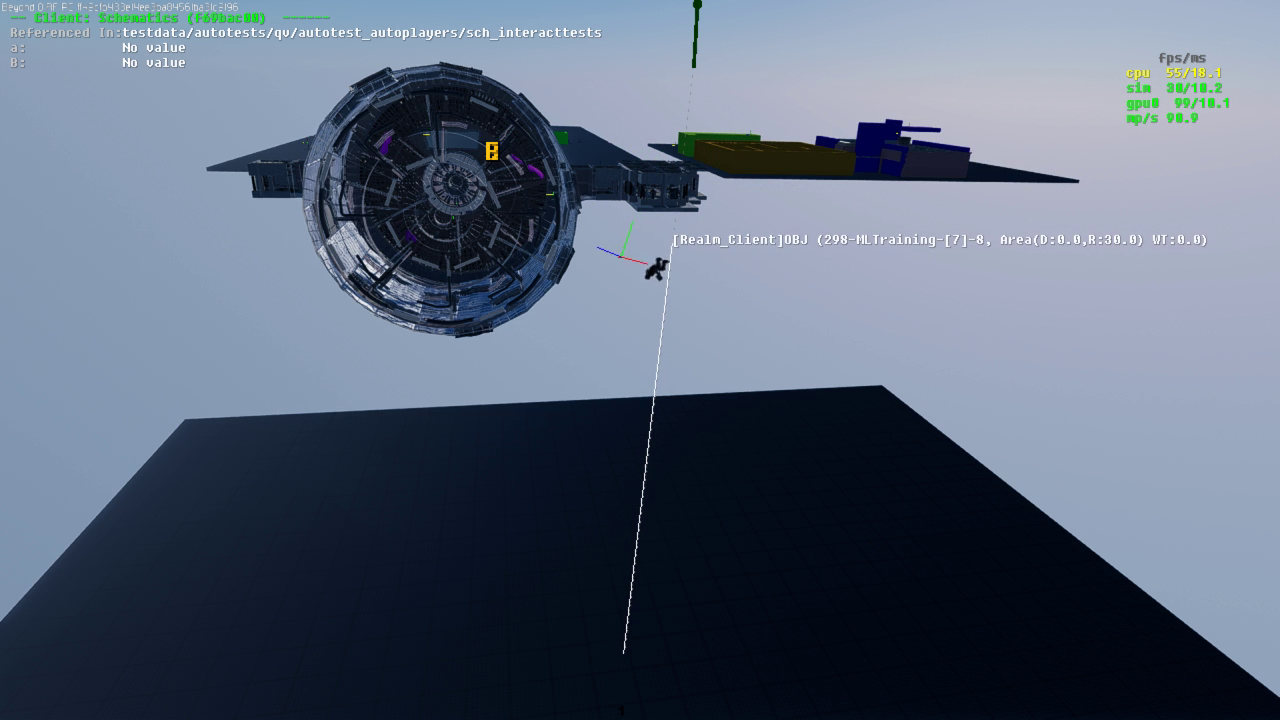}
        \caption{Test range environment}
        \label{fig:dead_space_test_range}
    \end{subfigure}
    \begin{subfigure}[b]{0.32\textwidth}  
        \centering 
        \includegraphics[width=\textwidth]{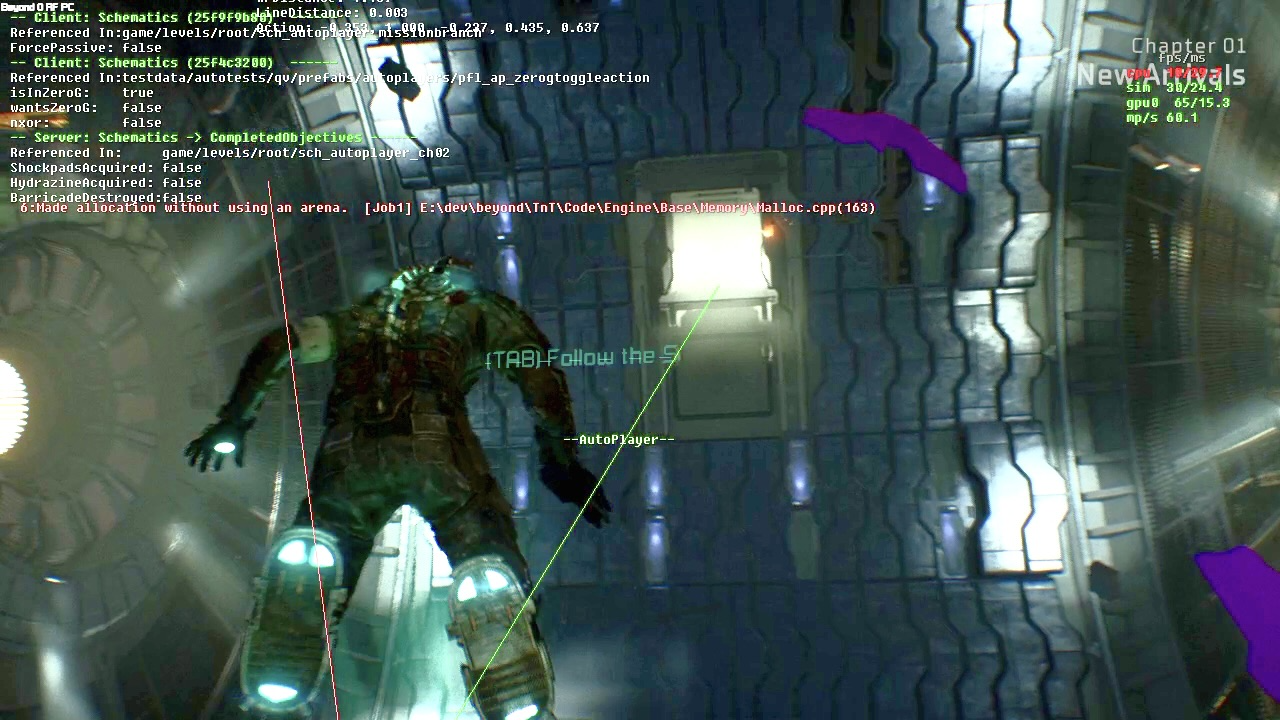}
        \caption{Production environment}
        \label{fig:dead_space_production}
    \end{subfigure}
    \caption{Examples of test range and full production environment of \textit{Dead Space (2023)}.}
    \label{fig:dead_space_environments}
\end{figure*}

Instead of training agents in the full production environments, test ranges were constructed containing only the essential features of the full game, reducing aspects of the game irrelevant to training. For instance, using a flat level with lower graphical fidelity by removing textures and collidable in-game objects enabled execution of substantially more agents in the same environment at a higher rate while still staying true to the full game using the same helicopter representation. A visual example of the differences between this test range and the actual production environment can be seen in Figure \ref{fig:test_range_helicopter}. As RL is sensitive to out-of-distribution observations, it should be noted that when using image data for training, textures and objects can not be removed as the visual interpretation of the game world would fundamentally change. A similar test range was constructed for \textit{Dead Space}, as seen in Figure \ref{fig:dead_space_test_range}, where the agent is tasked to navigate towards random target waypoints in a zero-gravity volume larger than the one experienced in the full production setting seen in Figure \ref{fig:dead_space_production}.

\subsection{Prototype environments}
\label{sec:prototype_environment}

\begin{figure*}[t]
    \centering
    \begin{subfigure}[b]{0.32\textwidth}
        \centering
        \includegraphics[width=\textwidth]{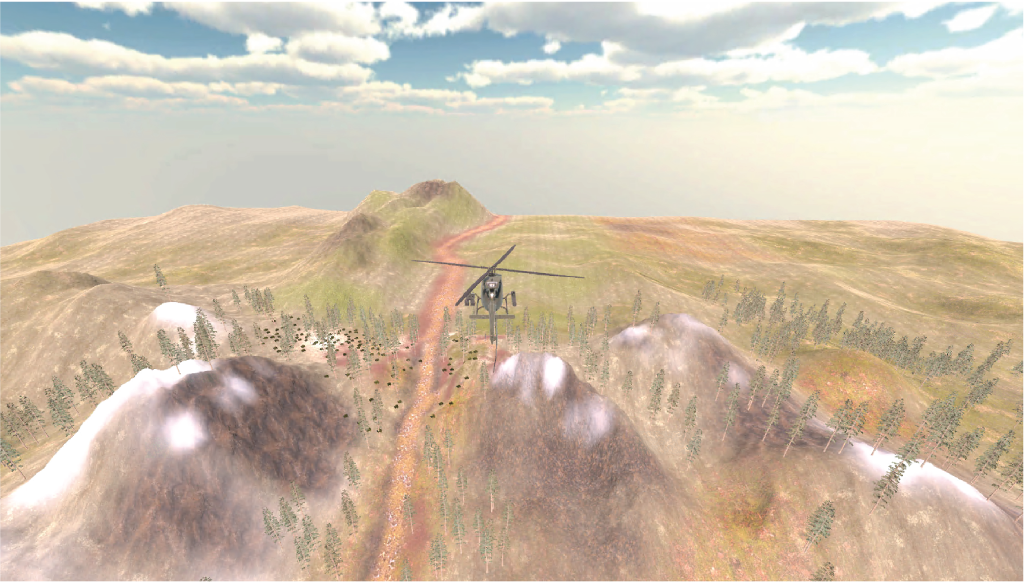}
        \caption{Prototype environment}
        \label{fig:prototype_helicopter}
    \end{subfigure}
    \begin{subfigure}[b]{0.32\textwidth}  
        \centering 
        \includegraphics[width=\textwidth]{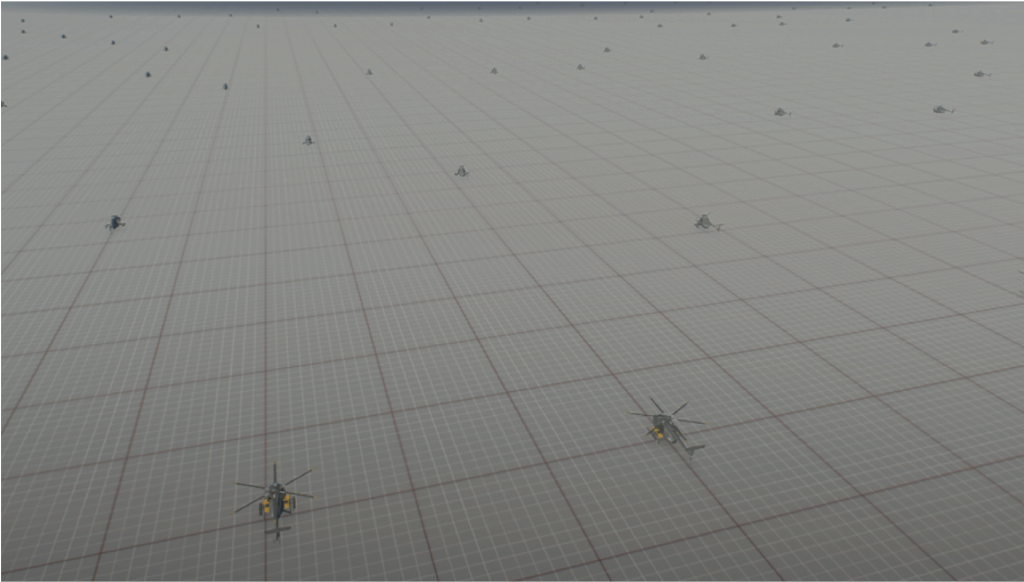}
        \caption{Test range environment}
        \label{fig:test_range_helicopter}
    \end{subfigure}
    \begin{subfigure}[b]{0.32\textwidth}
        \centering
        \includegraphics[width=\textwidth]{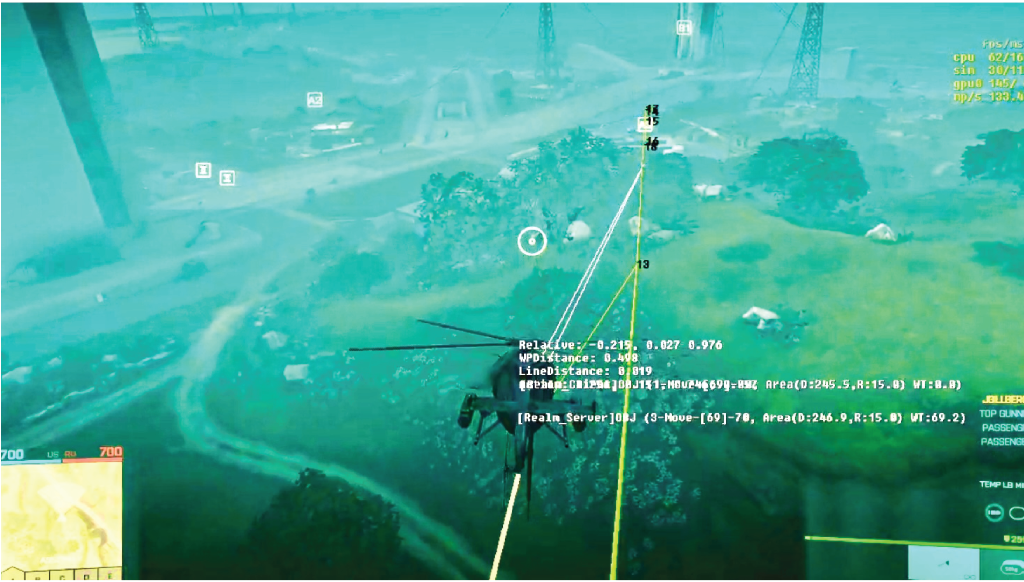}
        \caption{Production environment}
        \label{fig:kingston_helicopter}
    \end{subfigure}
    \caption{Example of environment setup going from prototype to production in \textit{Battlefield 2042}: three different RL training environments.}
    \label{fig:helicopter_environments}
\end{figure*}

There are significant benefits when training your agents in the same engine and game you are going to deploy them in. However, this is not always possible, especially early in production when the game is too unstable. To allow for easier setup prototyping, we created a simpler prototype environment in \textit{Unity3D} seen in Figure~\ref{fig:prototype_helicopter}, containing all key features that were expected from the final environments.

For example, if the task is to fly a helicopter, the prototype environment should implement the same actions, although without exactly the same flight control parameters like thrust power, torque, turning radius, etc. For RL, these discrepancies are learnable, which was proven both in theory and practice when we transitioned to the full environment. Similarly, for the observations, we used similar values available in the full environments from discussions with game team engineers. As both game scenarios involve 3D navigation, the setup tuning for the \textit{Battlefield 2042} case proved to be useful when also investigating \textit{Dead Space}, and the prototype environment allowed us to quickly iterate on observation representations, rewards and neural network architectures, ultimately leading to a reusable algorithm setup that needed very little tuning when deployed in the test range environments for both games.
\section{Reinforcement Learning for Game Testing}
\label{sec:method}

Inspired by research demonstrating how RL can be used to improve game state coverage \cite{bergdahl2020augmenting, gordillo2021improving}, we combined RL with \textit{AutoPlayers} bots to evaluate how efficiently the solution would be in a production setting for solving a real problem: game state coverage in hard-to-control vehicles and zero-gravity navigation. In this section, we describe and motivate the main elements of the RL setup used in this work.

\subsection{Augmenting bots capacity with Reinforcement Learning}
Given the navigation problems detailed in Section~\ref{sec:environments}, the idea was to replace the helicopter and the zero-gravity specific locomotion code with a version that instead infers movement controls using an RL model. This allowed the agent to easily switch between using \textit{AutoPlayers} locomotion code while on foot to RL based locomotion as soon as the agent entered the pilot seat of a helicopter or zero-gravity sections. This could also easily be extended by using different models for different vehicle types.


\minisection{Task Specification.}
In order to train an agent to control helicopters in \textit{Battlefield 2042}, the agent is initially placed on the ground inside the helicopter. From this position, we generate in-air target waypoints in a stepwise fashion requiring the agent to learn how to lift from the ground without crashing. The agent will then navigate between mid-air waypoints until ultimately arriving at a final waypoint situated at some location back on the ground-level, necessitating successfully landing. For \textit{Dead Space}, the navigation problem is made easier due to the lack of gravity. Hence, we task the agent to merely navigate between random waypoints in the play space.

\minisection{Parallelization Implications and Algorithm Choice.}
Given the parallelization constraints experienced when running complex games, low data collection rate inhibits the training process. There are different ways to mitigate this problem in practice. In general, leveraging distributed training, it is beneficial to run as many parallel environments as your hardware allows. Depending on the game, further optimization methods may or may not be possible. For games fundamentally designed around multiplayer such as \textit{Battlefield 2042}, the multiplayer concept can be reused to allow for the instantiation of multiple agents per launched game process as long as the model is able to produce actions for each agent at every inference step. In singleplayer oriented titles such as \textit{Dead Space}, this is, however, impossible without additional engineering. As such, the aforementioned techniques allowed us to run $250$ agents distributed over $5$ game servers simultaneously in \textit{Battlefield 2042} and merely $7$ in \textit{Dead Space}.

In this study, we evaluated a few algorithms with this in mind. On-policy methods such as Proximal Policy Optimization (PPO) rely on a continuous flow of fresh environment data, effectively discarding seen samples at each model update \cite{schulman2017ppo}. PPO is a popular on-policy algorithm due to its stability and lower tuning requirements. Inspired by the TRPO trust region policy optimization scheme, gradient backpropagation in PPO is bounded and never shifts the policy network too far from the current policy, leading to fewer uncontrolled updates. What makes PPO less viable is that, being an on-policy algorithm, it requires a massive number of samples before it starts to converge to a high-performing policy.

In addition to PPO, we also explored Soft Actor-Critic (SAC) which is an off-policy algorithm that stores data that come from different policies (e.g. past versions of the current model) for future model updates \cite{haarnoja2018soft}. From this, learnable information encoded in previous experiences can be reused, leading to higher sample efficiency. As a result, the number of environment interactions needed when using SAC is substantially lower than for on-policy algorithms. This, however, comes with a caveat: SAC tends to be unstable like the majority of off-policy algorithms~\cite{sutton2018reinforcement}. Even if the number of required environment interactions is reduced, the training policy is not guaranteed to converge to a good solution with high in-environment performance. As a result, it is possible the model needs to be retrained multiple times, lowering the gains from improved sample efficiency. In the end, even given the sample efficiency of SAC compared to PPO, the less stable results from SAC made PPO more favorable. 





\minisection{Observation Space.}
\label{sec:observation_space}
 In the case of \textit{Battlefield 2042}, simulating 250 helicopters over all game processes does not allow for efficient image based input generation as rendering viewports for all of them is intractable. Further, when using image data, the agent becomes sensitive to graphical changes in the environment, such as new textures, assets or rendering post processing common in the game production process. Hence, we decided to construct an observation space based on a minimal set of game state features. This state is ego-centric to the agent. Suppose that the world-space position of the agent at timestep $t$ is defined as $p_{t} \in \mathbb{R}^{3}$, then the state includes the relative positions of the current and next target waypoint $w_{t}^{i}, w_{t}^{i+1} \in \mathbb{R}^{3}$, distance to current waypoint $d_{w_{t}^{i}} = \| w_{t}^{i} \| \in \mathbb{R}$, closest distance to the line $l_{t}$ spanned by the vector from the previous waypoint to the current, defined as:
 \begin{equation}
 \label{eq:point_line_distance}
 d_{l_{t}}  = \frac{\|(p_t-w_{t}^{i-1}) \times (w_{t}^{i} - w_{t}^{i-1})\|}{\|w_{t}^{i}-w_{t}^{i-1}\|} \in \mathbb{R},
 \end{equation}
distance to ground $d_{g_{t}} \in \mathbb{R}$, velocity $\dot{p}_{t} \in \mathbb{R}^{3}$, acceleration $\ddot{p}_{t} \in \mathbb{R}^{3}$, quaternion rotation described by $q_{t} \in \mathbb{R}^{4}$, quaternion rotation difference $\Delta q_{t} = q_{t} - q_{t-1} \in \mathbb{R}^{4}$ since last rotation sample, and finally the $z$-axis angular alignment $\phi_{t} \in \mathbb{R}$ based on the dot product between the unit vector representing the forward direction of the agent and the normalized vector between the agent and current waypoint projected onto the $XY$-plane. All spatial values are normalized to a predefined bounding box encapsulating the entire game scene section that is playable to the agent. As the zero-gravity navigation problem specification in \textit{Dead Space} is similar to the helicopter case, the same observation space was used with the omission of height from the ground.

\minisection{Action Space.}
As the environments require fine control, we use continuous actions to allow the agent to make variable adjustments when navigating. Each action is represented by a vector $a_{t} = [a_{t}^{0}, ..., a_{t}^{N}]$, where $a_{t}^{i} \in [-1, 1]$. In the case of \textit{Battlefield 2042}, $5$ actions are used corresponding to throttle, pitch, yaw, roll, and a fire action. The latter is treated as a categorical action in the game side, which is triggered when the network output reaches a threshold value of $0.5$. For \textit{Dead Space}, we use a similar space of $5$ actions controlling pitch, yaw and roll, but with forward and strafe movement in place of throttle and the fire action.


\minisection{Reward Function.}
Due to the similarity between the tasks in each game, their corresponding reward functions are largely the same. We used the \textit{AutoPlayers} system to perform stepwise generation of valid target waypoints $w_{t}^{i} \in \mathbb{R}^{3}$ to construct a procedural navigation path described by the set \mbox{$\mathcal{W} = \{w_{t}^{0}, w_{t}^{1}, ..., w_{t}^{N-1}, w_{t}^{N}\}$} of connected waypoints starting from position $w_{t}^{0}$ to an ultimate goal location $w_{t}^{N}$. Note, that this sequence always starts and ends at the ground, requiring the agent to both lift off, navigate midair and finally land. We use the reciprocal closest orthogonal distance $d_{l}$ to the lines connecting consecutive waypoints following Equation \ref{eq:point_line_distance} to reward the agent for staying close to the shortest path between waypoints. To keep the agent facing the current waypoint, we add a reward based on $\phi \in [-1, 1]$. For each timestep, the agent gets an additional reward proportional to the difference between the last and current measured distance to the active waypoint $\delta _{w} = d_{w_{t-1}^{i}} - d_{w_{t}^{i}}$, which effectively penalizes the agent for moving away from the waypoint. Finally, the agent receives a reward for arriving at the current waypoint defined by $\mathds{1}(d_{w_{t}^{i}} < \epsilon)$, where $\mathds{1}(\cdot)$ returns 1 when its argument is true and 0 otherwise and $\epsilon$ defines the configurable specific threshold distance for arriving at the waypoint. The final reward is defined as:
\begin{equation}
    r_{t} = \frac{\alpha}{d_{l_{t}}} + \beta\phi_{t} + \gamma\delta_{w} + \psi\mathds{1}(d_{w_{t}^{i}} < \epsilon),
\end{equation}
where $\alpha$, $\beta$, $\gamma$ and $\psi$ are configurable reward shaping parameters. For the 
\textit{Battlefield 2042} use case, we require an additional reward term stemming from the fundamental difference between navigating with and without gravity, compounded by the fact that the helicopter is destroyed from collisions. As a result, we penalize the agent when the dot product between its upwards facing vector and the world-space unit vector $\hat{z}$ is $\leq 0$, which corresponds to the helicopter flying sideways or inverted, resulting in flight instability and a high likelihood of crashing into the ground. This phenomenon is especially detrimental when the agent is trying to land the helicopter at the final target waypoint on the ground.

\minisection{Model Architecture.}
Besides requiring expensive rendering operations, the use of image data also imposes a higher degree of model complexity as it relies on convolutional layers for feature extraction, which not only prolongs training but also increases inference times and sensitivity to visual changes in the environment. Following the previously defined observation representation, we used a multilayer perceptron for encoding states. The model was composed of two linear layers of sizes $512$ and $256$ for PPO and $512$ and $512$ for SAC, each layer using ReLU activations. Each model has an input layer of size $87$ corresponding to a 1D stack of $3$ observations of size $24$ for both games. Each model is trained to produce policies $\bm{\pi}(s; \theta) \sim \mathcal{N}_{|\mathcal{A}|}(\bm{\mu}, \bm{\Sigma})$ where $\bm{\mu}$ and $\bm{\Sigma}$ are learned, and correspond to the first and second moments of each of the $5$ continuous actions.

\begin{figure}[h!]           
	\centering				
    \includegraphics[width=1\columnwidth]{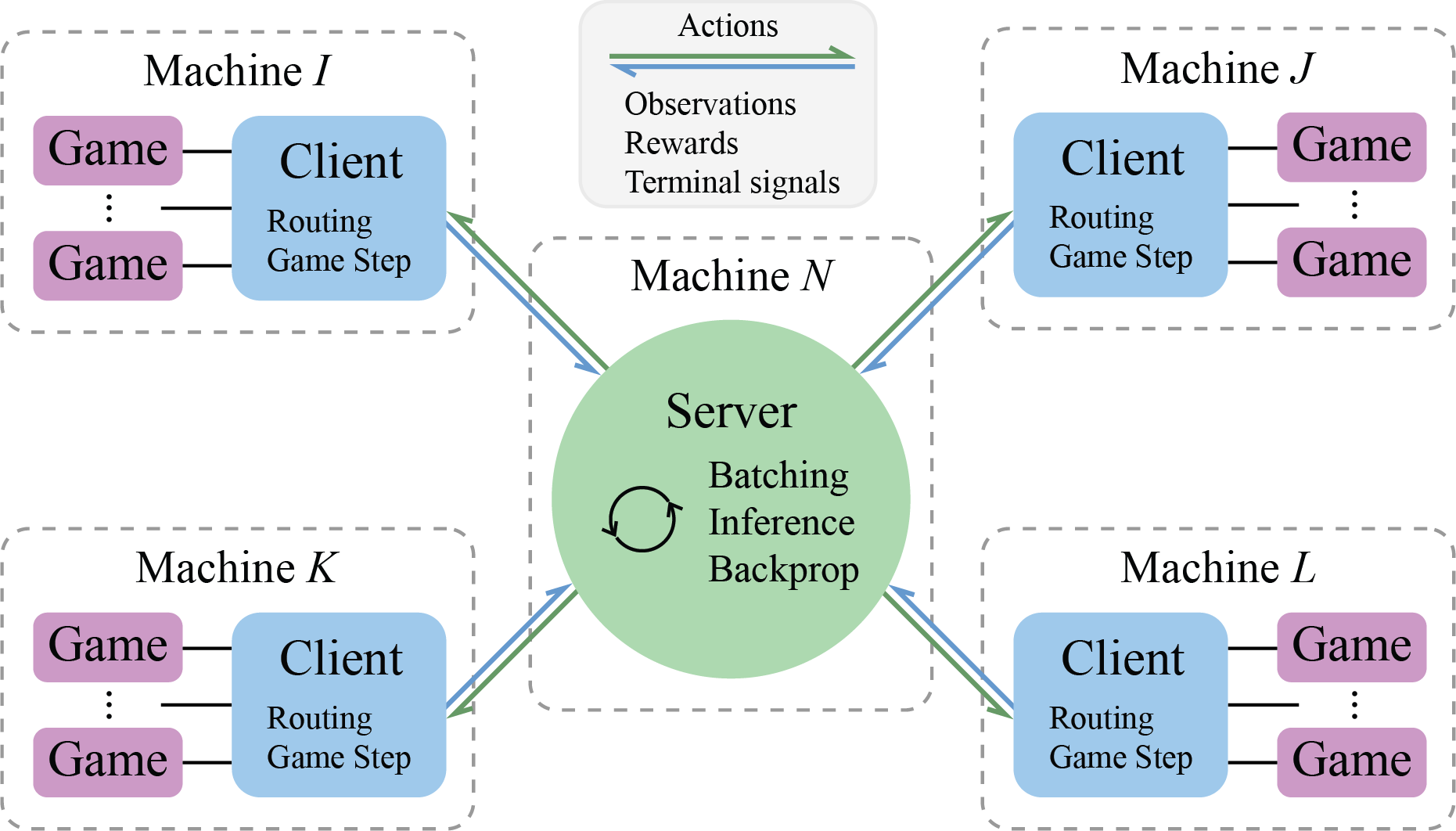}
    \caption{Architecture overview of the in-house RL framework used for training agents in this work. The framework distributes the collection of environment data over multiple clients, each in ownership of multiple game processes. The training server aggregates and batches data received from each client upon inference to update a central model. The client-server abstraction allows for the training to either be run locally or distributed over multiple machines, leveraging efficient communication over ZMQ network sockets.}
	\label{fig:seed-rl_architecture}	
\end{figure}

\minisection{Framework for training RL agents.}
\label{sec:seed-rl}
To train the PPO and SAC agents, we utilize an in-house trainer built around distribution and parallel data collection from multiple running game environments, illustrated in Figure \ref{fig:seed-rl_architecture}. The trainer makes use of a standardized OpenAI Gym-like API for stepping the environments, thus enabling support for a wide spectrum of platforms, including \textit{Frostbite}, \textit{Unity3D}, \textit{Unreal Engine}, and other classical OpenAI environments \cite{brockman2016openaigym}. Training is centralized following employing a client-server network topology where multiple clients are served action data upon inference from a server that hosts the neural network model. The clients return observation and reward data from each game, collated following unique IDs reflecting each client-hosted game process, allowing the server to batch data used to perform the gradient backpropagation needed to update the model. Further, the trainer produces models supported by the Open Neural Network Exchange (ONNX) standard~\cite{bai2019onnx}, allowing for embedded inference with the ONNX runtime~\cite{onnxruntime}. Having a general, centralized training platform allows us to easily share models and training setups across teams. 

By abstracting away the complexity of machine learning algorithms, this modular library of building blocks enables rapid development and iteration. Further, by distributing data collection across many processes and machines, it allows for a high-performing training procedure, lessening training time to allow for faster iteration. To allow for training in the mentioned production and test range environments, an API between the internal training framework and \textit{AutoPlayers} was set up to enable the communication of actions, rewards and observations to and from the \textit{AutoPlayers} player-endpoint.

\subsection{Deployment}
To use the trained RL models in-game, runtime support was needed since there was no guarantee that the game clients always would have access to the trainer for all potential environments and setups the agents could be used. This is especially true when running training on various platforms and cloud solutions. The exported ONNX-model was embedded and baked with the rest of the game data and loaded by the helicopter locomotion code when the agent enters a vehicle. This ML locomotion collects observations in the same way, using the same code, as was done during training by reading values from the \textit{AutoPlayers} API and then feeding them to the ONNX runtime. The output of the network is then interpreted as movement controls and injected as such together with other AutoPlayer generated inputs.

Generally, inference support in game engines is relatively new and improving rapidly with respect to performance and compatibility with other game systems. Integration with other systems frequently requires operating on a batch size of 1 which means that batching for speed-ups in inference is not always possible. Inference run-time is critical for production, where current solutions for automated testing only require \SI{<100}{\us} for decision making, setting an upper feasibility boundary for inference time. The ONNX runtime provides a good starting point, but for AAA games that would like to use ML models for many features, or for many agents, a bespoke high-performance inference library will probably be needed. In our experiments, we were limited to fairly small models due to previously mentioned restrictions ranging from $10^{3}$ to $10^{6}$ parameters in size. As inference support improves, these models could be increased in complexity. Furthermore, quantization of model parameters is a good way of increasing performance and reducing memory footprint but it was not used in our experiments.

\section{Lessons Learned and Discussion}
\label{sec:discussion}
In general, RL integration comes with a set of requirements on the game often not seen in other methods, and these need to be fulfilled. Identify test cases early where RL could provide value, as the implementation procedure of the method comes with issues that might be easy to solve early in game production but virtually impossible at a later stage. In this section, we list formulated suggestions and lessons learned during this project. This list is not exhaustive, but it can provide a starting point for preparing your game for RL integration and bring awareness to potential causes of sub-optimal training performance you may encounter. 

\begin{itemize}
\item Use the most comprehensive and meaningful observation representation up until simulation performance is degraded. For AAA games, every microsecond of CPU time matters. In many cases, this means that computationally expensive ways to collect data about the world such as raycast queries or vision data rendering can be prohibitively costly. Remember that data has to be collected not only for the training procedure, but also for when the in-game agent needs to use the model for inference at runtime. In addition, try to use the smallest possible model that does not adversely affect agent performance.

\item Start from the most stable version of the game, branch from that, and only start accepting upstream updates once successful training has been established. This is especially important while iterating on the training setup, thus it is essential to have a plan for how to provide a stable training environment. Games under development are not stable, and  both keeping up with and training agents directly in production code is virtually impossible.

\item Establish a plan for keeping the training setup alive by continuously testing it. Even if a model is successfully trained early in development, the game will keep changing. As the game changes, it is almost certain that the model will have to be retrained at some point. This entails tuning hyperparameters, observation spaces and rewards, or recording new data in the case of Imitation Learning (IL) or just ascertaining that your training setup still works.

\item As the game development process rapidly advances, make a decision on what method to use beforehand as there may be no time to fully explore alternatives. Given the nature of the studied environments in this work, as IL generally requires less training time it could be a sensible alternative to RL. We made initial explorations using Generative Adversarial Imitation Learning (GAIL) in \textit{Battlefield 2042}, but given the state of the game, demonstration recording was difficult as the helicopters were still too hard to control and further user-facing improvements were yet to be implemented \cite{ho2016gail}. Additionally, any changes made to the game would make any already recorded demonstration data obsolete.

\item Leverage the knowledge and efficiency of scripted solutions and only use RL when needed. We believe that the ideal approach is to complement the automated scripting system with RL, rather than use a full end-to-end RL solution to utilize existing knowledge and test procedures. There are many test cases that require a level of control best achieved through scripting, so it is more efficient to put such a system in place first before RL is applied to improve certain parts of it.

\item Create your observation space with care. It is important to ensure that every reward signal has an appropriate number of observations to which it can be assigned. Having too many observations is preferable to having too few. RL requires a solid environment definition and is particularly sensitive to poorly defined observation-spaces and reward signals, and their relation to each other: any \textit{unexplained} reward will significantly reduce training efficiency. For instance, if a helicopter gets a negative reward when colliding with the ground, make sure an observation that either measure the distance to the ground, or any other indicator exists otherwise training will be undone if a seemingly random reward is propagated through the network.

\item Make sure you are able to scale your game by running many instances of the game in parallel, and/or running many agents in each game process. Further, fast forwarding the game at a more than 1x simulation rate should allow for faster training. This requires that the machine is actually capable of running the game faster but even then the physics engine within the game is prone to failing, especially for validating collisions. When this was attempted for the helicopter scenario, the parked helicopters would fall through the ground at the start of the training session. In general, RL requires many environment samples and choosing an algorithm for training the RL agent has to be done with consideration. Off-policy algorithms such as SAC can be more sample efficient than on-policy methods like PPO, but with the trade-off of being more unstable \cite{sutton2018reinforcement, haarnoja2018soft}.

\end{itemize}

\section{Related work}
\label{sec:related_work}
The goal of this paper is to provide guidance to engineers when applying RL to game testing in a production environment. The potential of deep reinforcement learning for video game testing has been gaining interest from both the research and video game communities. Here, we review work from the literature most related to our paper.

\subsection{Game testing}
Recently, automated playtesting and validation techniques have been proposed as a way to reduce manual validation in large games. These works have primarily relied on classical hand-scripted AI~\citep{stahlke2020pathos,holmg2018personas,chang2019reveal} or proposed model-based testing, such as ~\citet{ftikhar2015platform} for platform games, and~\citet{lovreto2018mobiletest} for mobile games~\citep{stahlke2019playfulness, schaefer2013crushinator, cho2010scenariobased, xiao2005fifa}. The work of \citet{mugrai2019automated} is particularly noteworthy, as it developed an algorithm for mimicking human behavior in an effort to improve the game design process as well as the validity of playtesting results. However, when dealing with complex 3D environments these scripted or model-based techniques are not readily applicable due to the high-dimensional state space involved and poor generalization performance.

\subsection{Machine learning and game testing}
Related work on real-world applications regarding controlling autonomous helicopters with RL and expert data has previously been studied in \citet{abbeel2010autonomous}. Recent research includes the use of RL to augment automated playtesting~\cite{bergdahl2020augmenting}. \citet{alonso2021navigation} trained an RL agent to navigate a complex 3D environment toward procedurally generated goals, while \citet{devlin2021navigation} trained different agents and classification models to perform a Turing test to evaluate the human-likeness of trained bots. \citet{zheng2019wuji} proposed an on-the-fly game testing framework that leverages evolutionary algorithms, DRL, and multi-objective optimization to perform automated game testing. \citet{agarwal2020} trained RL agents to perform automated playtesting in 2D side-scrolling games along with visualizations for level design analysis, and \citet{gordillo2021improving} used intrinsic motivation to train many agents to explore a 3D scenario with the aim of finding issues and oversights. Similar to the latter, \citet{sestini2022ccpt} proposed the curiosity-conditioned proximal trajectories algorithm with which they test complex 3D game scenes with a combination of IL, RL, and curiosity driven exploration. Nevertheless, the majority of the previously cited approaches were used in proof-of-concept games and, despite their complexity, still provide a simplistic view of applying RL methods to an actual game in production.

\subsection{RL in game production}
At the same time, there are only a few examples in the literature regarding RL in game production. Notable examples are: \citet{harmer2018imitation} proposed a multi-action imitation learning framework used in the game \textit{Battlefield 1}; \citet{iskander2020reinforcement} trained multiple RL agents to play the multiplayer game \textit{Roller Champions}; \citet{wei2022honor} train a single complex agent in one of the most popular game \textit{Honor of Kings}; finally, GTSophy~\cite{wurman2022outracing}, an RL policy trained to beat professional players in the game \textit{Gran Turismo 7}, was recently shipped within the actual video game. Despite the already cited papers discussing RL, most do not address the challenges that a game studio must face when applying these solutions for either validating their games or creating non-player characters. 

\section{Conclusion and Future Work}
Herein we have described a cross-disciplinary and cross-team experiment to bring a research prototype into AAA game production. The main finding of this paper is that indeed it is possible, although difficult, to bring RL into an existing automated and scripted bot system. We have shown how RL agents can add capability to an already capable automated testing system.  Delivering RL capacity into the hands of QV testers was achieved through extensive collaboration between researchers, QV and the game teams. 

Machine learning offers a new way of creating behaviors that can be used for testing. By talking to Quality Verification (QV), Quality Assurance (QA), and other experts in related fields, we were able to gather a lot of potential future work in that area. It can be divided into two different needs: the learning capacity of the algorithms, and the technical aspects. In areas of the model learning capacity, we see some important improvements that most likely will require advancements in machine learning to achieve. We also see that there are hurdles when having non-experts using ML. In most literature, part of the impressive performance and results comes from the fact that it has been trained and set up by a field expert and it might not even be possible to otherwise do so without that person. Overall, in order for ML/RL to be used by laymen some usability improvements in algorithms would be desired. Here we list a few topics that we think would help increase adaptation in games and game production.

\begin{itemize}
\item More stable and predictable RL algorithms: in a production setting there will be little time for parameter tuning, observation tuning, etc. It is desirable to have a robust algorithm that arrives at a stable solution in many different game iterations at a slower pace rather than to have a faster but unstable algorithm that provides high variance in performance which requires more training iterations per version of the game.
\item Behaviour explainability: understanding a behavior, in order to correct it, is essential. When an RL agent gets confused and ends up stuck in a critical place of the game due to a bug, without explanation it would be hard to correct said bug.
\item Reward shaping: more intuitive ways of doing reward shaping. Most game developers lack the RL/ML expertise needed to write good reward functions for specific testing behaviors.
\item Transfer learning: games in production are constantly shifting, and improving how a model adapts to a new game environment would significantly decrease re-training needs and speed up the process.
\end{itemize}

Many of these problems are fundamental to machine learning in general, but hopefully this will give an indicator of where we see improvements can be made and what research is needed for adaptation to game production. Games and game production is a field that could heavily benefit from more user-friendly and capable machine learning algorithms.


{\footnotesize \bibliography{main}}
\bibliographystyle{IEEEtranN}

\end{document}